\documentstyle[prl,aps,multicol]{revtex}
  \renewcommand{\narrowtext}{\begin{multicols}{2} \global\columnwidth20.5pc}
   \renewcommand{\widetext}{\end{multicols} \global\columnwidth42.5pc}

\begin{document}

\title
{Acoustoelectric effects in quantum constrictions}
\author{Frank A. Maa\o$^1$ and Y. Galperin$^{2,3}$} 
\address{$^1$ Institutt for fysikk, The Norwegian University of
Science and Technology, 
NTNU, N--7034 Trondheim, Norway \\
$^2$ Fysisk institutt, University of Oslo, P. O. Box 1048 N--0316
Oslo, Norway\\ 
$^3$ A. F. Ioffe Physico-Technical Institute R A S, 194021
St. Petersburg, Russia} 
\date{\today}
\maketitle

\begin{abstract}
 A dc current induced in a quantum constriction by a traveling
 acoustic wave (or by non-equilibrium ballistic phonons) is
 considered. We show that in many important situations the effect is
 originated from acoustically-induced scattering between the
 propagating and reflecting states in the constriction. Two particular
 regimes corresponding to relatively high and low acoustic frequencies
 are discussed. In the first regime, the acoustoelectric effect in
 a smooth constriction can be understood by semi-classical
 considerations based on local conservation laws. For the low
 frequency regime, we show that the acousto-conductance is 
 closely related to the zero field conductance.
The qualitative considerations are confirmed by numerical calculations
 both for smooth and abrupt channels.
\end{abstract} 

\narrowtext

\section{Introduction}\label{Introduction}
Recently surface acoustic waves (SAW)\cite{Shilton1,Shilton2,Nash96},
as well as non-equilibrium ballistic phonons\cite{Naylor}, have been
successfully used to study mesoscopic systems with a
quasi-one-dimensional electron gas as quantum point contacts (QPC)
and quantum wires (QW). In most cases these systems connect two
reservoirs made of a two-dimensional electron gas (2DEG). Consequently,
these studies were a natural continuation of an extensive work on
acoustical studies of 2DEG. 

Due to the piezoelectric effect, acoustic waves induce traveling waves of
electric field which propagate with the sound velocity $s$. Being
concentrated near the surface, these waves are extremely sensitive to
the electric properties of the relevant region. In particular, a highly
conductive 2DEG can screen out the electric fields and in that way
influence the sound velocity. If the electric field is not screened
completely, the Joule losses in the 2DEG lead to a pronounced
attenuation of the waves. In both situations acoustic waves provide a
possibility of probe-less measurements of the 2D conductance
$G(\omega,\bf{q})$ as a function of the sound frequency $\omega$ and
wave vector ${\bf q}$. The role of screening is determined by the
dimensionless ratio $G/s$ (note that $G$ has the dimensionality of the
velocity in the CGSE system) which is usually large (see
e.g. Ref.~\onlinecite{EG}). However, the 
component of $G$ parallel to the surface can be suppressed by a magnetic
field and thus made of the same order as the sound velocity. This is
the case, in particular, under the conditions of the quantum Hall effect,
and several beautiful and instructive experiments were conducted in
that regime\cite{QHE1,QHE2}. Many important experiments were performed
using ballistic phonons (see, e.g., Ref.~\onlinecite{challis}) which
can be induced by laser beams\cite{Karl} or by
resistive film heaters~\cite{Naylor}.

Two groups of experiments are usually conducted. In the experiments of
the first group, a {\em linear} response to the wave is
studied. Namely, the variations in the sound velocity and attenuation
are registered. The second group of 
experiments\cite{Shilton1,Shilton2,QHE2} is concentrated on the
studies of the response of the electron gas to the acoustic waves. The
quantities which are investigated in these experiments are the dc
electric current induced by the waves (the so-called {\em
acoustoelectric current}), or the dc voltage induced across the
sample. These effects are {\em non-linear} in the sound amplitude, at
small amplitudes they are proportional to the amplitude squared.  

The acoustic waves seem very advantageous for studies of quantum
constrictions like QPC or QW. Indeed, the screening length in the 2DEG
coincides with the Bohr radius, $a_B$. Consequently, if the dimensions of
a 2D sample are much larger than $a_B$ then the (piezoelectric)
coupling to phonons is screened out, at least in the absence of a
perpendicular magnetic field. However, if the width of the channel is not
exceeding $a_B$ by too much, the field penetrates the channel. As a
result, it is the channel region that contributes to the interaction
rather than the 2D leads where the coupling is screened out. On the
other hand, the resistance of the system is also determined by the
channel, and it is possible either to monitor its changes under the
influence of the waves (the so-called acousto-conductance), or to
measure the acoustoelectric current.\cite{end0}

 Recently, two experiments \cite{Shilton1,Shilton2} on the induced
acoustoelectric current in quantum constrictions revealed several interesting
features. In the first paper by Shilton et al.~\cite{Shilton1} the regime of
weak acoustoelectric current was investigated, while the second
paper~\cite{Shilton2} examined the regime of a strong SAW-field. In
the large-amplitude regime 
the induced current was shown to be quantized in units of $e f$, with
$e$ being the electronic charge and $f=\omega/2\pi$ the wave frequency. This
behavior was explained by the authors of
Ref.~\onlinecite{Shilton2} in terms of `traveling quantum dots'
with 
an integer number of electrons in each dot. Thus, during one period an
integer number of electrons are carried from one side of the constriction
to the other. On the other hand, the weak-field regime displayed a 
complicated pattern of peaks and dips close to pinch-off and a more regular
pattern of oscillations for higher gate voltages. It was noted that
the structure of the acoustoelectric current resembled (but did not
coincide with) that of the derivative of the zero field conductance
with respect to the gate voltage. 

The low-amplitude behavior of the acoustoelectric current was
discussed in Ref.~\onlinecite{Totland} on the basis of the Boltzmann
equation for the electrons interacting with the traveling
wave. The theory, which was able to reproduce the main features of the
experimental observations, is based on a model treatment of the
finite channel's length and of the role of elastic scattering. A
quantum theory of the acoustoelectric effect in a uniform quantum
channel was developed in Ref.~\onlinecite{Gurevich} in an analogy to the
previous treatment of equilibrium phonons\cite{Gurevich1}. The main
point was to analyze the energy and momentum conservation laws which
correspond to backscattering of the electrons by the traveling wave.
It was shown that there is a cut-off in the electron-SAW interaction
at $q = q_0 \equiv 2ms/\hbar$. At $q \le q_0$ backscattering is
forbidden, while the forward scattering cannot influence the current
in a 1D system. This statement has been also expressed in a recent
paper\cite{Blencowe} where the interaction with non-equilibrium
ballistic phonons was considered. At the same time, in the experiments
\cite{Shilton1,Shilton2} the interaction is clearly observed despite
of the inequality $q \ll q_0$.

Our aim is to draw attention to the fact that in any realistic quantum
channel, there are special regions near the leads where an acoustic
wave (or
non-equilibrium ballistic phonons) can cause transitions between the
{\em propagating} states in the channel and the {\em reflecting} states
which correspond to reflection of an electron back into the incident
lead. These 
particular processes facilitate phonon-induced backscattering. As
a result, the acoustoelectric effect persists at $q \le q_0$ and
several new and important features of the effect appear at $q \ge
q_0$. Note that the role of the processes mentioned above was
emphasized in Ref.~\onlinecite{Grincwajg} in connection with
photo-conductance of quantum channels. However, there are  important
differences between the case of transverse electro-magnetic field with
zero wave vector (as in the case of photo-conductance) and the longitudinal  
wave with a finite wave vector (as in the case of the acoustoelectric
effect). Below we present a quantum theory which takes into
account the above mentioned transitions and in this way explains the
results obtained in realistic structures. We believe that at $q \ll
q_0$ these transitions lead to the same physical effect as the
introduction of a finite relaxation length in the classical
theory\cite{Totland}, while they do not have analogs in the theory for
an uniform channel\cite{Gurevich,Blencowe}.

The paper is organized as follows. In section \ref{Theory} we briefly
go through the basic assumptions and formalism. Section \ref{HF} is devoted
to simple semi-classical analysis and is most relevant for high frequency
fields as generated with the focused laser beam techniques. Section \ref{LF}
concerns the low frequency regime where semi-classical theory is invalid.
This section is relevant for frequencies generated by SAW techniques.


\section{Theory}\label{Theory}

Being interested in the response of quasi-one-dimensional systems 
(quantum constrictions, wires, etc.) we assume that the channel's width is of the 
order of the Bohr radius $a_B$ while the leads consist of a two-dimensional electron 
gas and have widths $\gg a_B$. On the other hand, the typical wave lengths $2\pi/q$ 
of SAW or non-equilibrium phonons is $\ge a_B$. As a result, the (piezo)electric 
field produced by acoustic waves is efficiently screened outside the channel. 
This is the reason why the 
response of a short channel is not completely masked by the leads. Consequently, 
the field is taken into account only in the region near the constriction.
Apart from this, we assume that the electrons are non-interacting and spin degenerate. 
For simplicity we consider the case of zero temperature and assume the quantum channel 
to be symmetric with respect to reflection of the longitudinal coordinate 
($x \rightarrow -x$). 
Under such an assumption there is no so-called ``rectified current", and the total dc
current is due to directed propagation of the acoustic waves. In a real experimental 
situation there is no full symmetry of the channel. However the rectified current can 
be subtracted by taking the average of the results obtained by reversing the direction 
of SAW (or phonon) propagation. 
Let the channel be elongated along the $\bf x$-axis, its dimensions in the $\bf y$- and
$\bf z$-directions being much less than $q_y^{-1}$ and $q_z^{-1}$, respectively. 
In that case non-equilibrium phonons cannot produce transitions between the modes of
transversal electron motion. Consequently, all transverse modes are
subjected to the sliding potential 
\begin{equation}
V(x,t) = V(x)\cos (qx-\omega t), \quad q \equiv q_x \, ,
\label{pot1}
\end{equation}
since under the assumptions made above the envelope function $V(x)$ is independent
of the transverse mode number\cite{end1}.

Neglecting spontaneous emission in the constriction region, we can write the dc 
electric current from left (L) to right (R) as \cite{Datta,Maao1} 
\begin{eqnarray} \label{cur01}
I & = & \frac{2e}{h} \sum_n \int_{0}^{\infty} 
\left[T_{R,L}(E+n\hbar\omega,E) f_{\mu_L}(E)- \right. \nonumber \\
& & \mbox{\hspace{1.7cm}} \left.
T_{L,R}(E+n\hbar\omega,E)f_{\mu_R}(E) \right] \, \mbox{d}E\, .
\label{current1}
\end{eqnarray}
Here $e$ is the electronic charge while
$T_{R,L}(E+n\hbar\omega,E)$ is the sum (over the transverse modes) of
transmission probabilities for particles 
with the energy $E$ to be transfered  {\em from the left to the right lead} into 
states with 
the  final energy $E+n\hbar\omega$.  $T_{L,R}(E+n\hbar\omega,E)$ denotes the reverse
transition, both  transitions being accompanied by absorption or (induced) emission 
of $n$ phonons. Eq.~(\ref{cur01}) is actually a generalization of the
Landauer-B\"uttiker formula\cite{Landbutt}.

Since we are going to study dc response in the absence of external bias, i.e. at 
$\mu_L = \mu_R = \mu$, the current can be expressed as 
\begin{equation}
I(\mu)  =  \frac{2e}{h} \int_{0}^{\infty}
\Delta T(E) f_{\mu}(E) \, \mbox{d}E,
\label{current2}
\end{equation}
where
\begin{equation}
\Delta T = \sum_n \left[
T_{R,L}(E+n\hbar\omega,E) - T_{L,R}(E+n\hbar\omega,E)\right].
\label{dT_def}
\end{equation}
A positive $\Delta T(E)$ means that non-equilibrium phonons enhance the 
left$\rightarrow$right ($L\rightarrow R$) dc particle current.  It will be 
illustrative  later to analyze the
behavior of $\Delta T(E)$ in terms of partial changes in the $L\rightarrow R$ and 
$R\rightarrow L$ 
transmission probabilities. We therefore define
\begin{eqnarray}
\Delta T_{R,L}(E) & = & \sum_n T_{R,L}(E+n\hbar\omega,E) - T^0(E)\\
\Delta T_{L,R}(E) & = & \sum_n T_{L,R}(E+n\hbar\omega,E) - T^0(E),
\end{eqnarray}
where $T^0(E)$ is the sum of transmission probabilities in the absence of acoustic 
perturbation.

In this work we shall study the regime of relatively small acoustic amplitudes where 
a perturbative approach is adequate. In particular, we restrict ourselves with the 
{\em linear in acoustic intensity} approximation where
it is sufficient to consider $n=-1,0$ and $1$. In the following we present the 
numerical analysis of the induced current based 
on the  recursive Green function method near the continuum limit \cite{Maao1}.
This method gives an exact solution to the scattering problem posed by the
Schr\"odinger equation. For small systems, such as those under the
present consideration, super-computing facilities are not necessary.
If not otherwise stated, the quantum channel consists of 2DEG, and its shape is chosen 
in the form 
\begin{equation}
\frac{W(x)}{W_{\infty}} = 
\frac{W_0}{(W_{\infty}-W_0)\cos^4(\pi x/2L) + W_0}\, .
\label{shape}
\end{equation}
Here  $W_{\infty}$ is the width in the region of the leads, $W_0$ is the minimal 
width, while $L$ is the characteristic length. The envelope function, $V(x)$, in 
equation (\ref{pot1}) is chosen as
\begin{equation}
V(x) = V_0 [1-\sin^8(\pi x/4L) ],
\end{equation}
when $|x| < 2L$ and zero otherwise.
The square-box confining potential
in the $\bf y$-direction is chosen such that the transverse wave functions vanish 
at $y=\pm W(x)/2$.
For the calculations presented we have chosen
$W_\infty = 8 W_0\simeq 0.3 \ \mu$m and $k_F\simeq 1.25\times10^8$ m$^{-1}$.
The sound velocity is set to $s$ = 2700 m/s, and the effective mass 
$m^*$ = $0.067 m_e$, typical for AlGaAs-structures.

The general behavior of $\Delta T(E)$ appears rather complicated. However, as it will 
be shown below, there are limiting cases of low and high frequencies where the basic 
features can be understood from very simple arguments.
We give these arguments below.

\section{High frequency regime}\label{HF}

In the high-frequency regime it is natural to consider the acoustically induced 
effects as scattering of electrons by non-equilibrium phonons. For simplicity,
let us consider a constriction with smooth variations
on the length scale set by the Fermi wavelength $k_F^{-1}$.
In this case we can use an {\em adiabatic} and {\em semi-classical} 
analysis\cite{glks}.
The essence of the {\em adiabatic} approximation is to
assume the electronic wave function can be expressed as a product of the 
{\em transverse} part (which varies slowly as a function of the longitudinal 
co-ordinate $x$), and the {\em longitudinal} part which bears the main $x$-dependence.  
Consequently, the shape of the constriction manifests itself only as rather smooth 
$x$-dependence of the transverse eigen energies, $E_m (x)$, for different transverse 
modes.  This $x$-dependent eigen energy, $E_m(x)$, will now act as a
potential barrier for an one-dimensional scattering problem describing electron 
transfer in the $x$-direction.
We put the origin of the reference frame in $x$-direction at  the maximum of $E_m(x)$ 
where the constriction is the narrowest.  

The {\em semi-classical} approximation ignores
quantum mechanical tunneling. This implies that the zero field transmission
probability, $T^0(E)$ for the $m$th mode, is zero when $E<E_m(0)$ (the
reflecting region) and one when $E>E_m(0)$ (the propagating region).
The range of applicability of the semi-classical approximation is
determined by the energy window, $\delta E$, where $T^0(E)$ goes from zero
to one. A sufficient criterion is $\hbar \omega \gg \delta E$. 

An important feature of the semi-classical situation is that the phonon-assisted
processes are {\em spatially confined} to regions relatively small in comparison 
to the total length of the channel, $L$. Indeed, in complete analogy with the 
semi-classical theory of photo-conductance \cite{Grincwajg}, only in the 
vicinities of the points $x^*$, corresponding to local conservation for both 
momentum and energy, the phase of the integrand which enters the perturbation matrix
element $\langle i|V(x,t)|f\rangle$ is a slow function of $x$. Consequently, only the 
vicinities of such points are important,  and the general behavior of the induced 
current can be understood from the analysis of the local conservation laws.

In the semi-classical approximation, the longitudinal part of the wave function 
can be expressed as
\begin{equation} \label{f11}
\psi_\parallel =\sqrt{\frac{k(\infty)}{k(x)}}
\left\{ \begin{array} {ll}
\exp \left(\imath \int^x k(x') \, dx \right),& E > E_m (0)  \\
\sin \left(\int^x k(x') \, dx \right),& E < E_m (0) 
\end{array} \right.
\end{equation}
where the local longitudinal momentum is defined as 
\begin{equation} \label {k1}
\hbar k(x,E) = \sqrt{2m^*(E-E_m(x))}\, .
\end{equation}
{}From now on, assume that the sliding potential in equation
(\ref{pot1}) is moving from the left to the right (i.e. $q > 0$).
 Conservation of energy and local longitudinal momentum implies
that intra-band transitions can only take place between two
states (labeled 1 and 2) at points $x^*$ when
\begin{equation}
k_1 (x^*,E) =  \frac{q_0 - q} {2} \, , \  
k_2 (x^*,E+ \hbar \omega) =  \frac{q_0 + q}{2}\, , \label{k2}
\end{equation}
where $q_0 \equiv 2ms/\hbar$, $m$ is the effective mass, $s$ is the
sound velocity.  
Here we have taken into account that for acoustical phonons $\omega = sq$.
Note that the component $\propto \exp(iqx-i\omega t)$ in $V(x,t)$ will
cause a transition from the state 1 to the state 2 (absorption), while
the component $\propto \exp(i\omega t-iqx)$ will cause a transition
from the state 2 to the state 1 (emission).
One can observe the characteristic quantity $q_0$ which sets the scale
for the phonon wave vector $q$. 
\begin{figure}[htb]
\vspace*{8cm}
\includegraphics{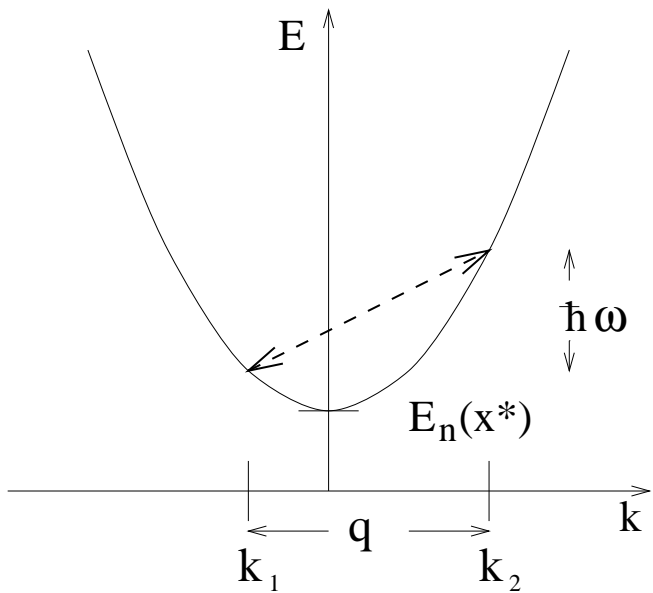}
\caption{Transitions that satisfy both conservation of momentum and energy}
\label{dispersionfig}
\end{figure}

Consider a long channel with an uniform width such as was considered
in\cite{Gurevich,Blencowe}. In such a channel, only scattering
between states {\em propagating} in opposite directions can contribute 
both to the drag current and to the acousto-conductance. 
This implies that we must have $q > q_0$ in order to see any effect.
However, for non-uniform wires this is no longer the case.
As it follows from Fig.~\ref{dispersionfig} and
Eqs.~(\ref{k2}), it is natural  to analyze the cases $q < q_0$ and $q > q_0$ 
separately.

\subsubsection*{The case $q < q_0$}

The case $q < q_0$ is the simplest. Although the criterion $\hbar
\omega \gg \delta E$ for the semi-classical treatment is usually not
met at $q < q_0$ in realistic systems, this approach is very
illustrative and leads 
to qualitatively correct answers. This is why it  has recently been
used to study the acoustoelectric current  
\cite{Gurevich} and acousto-conductivity \cite{Blencowe} in quantum wires. 
The authors concluded that neither current nor the change in the
conductance should be induced at $ q < q_0 $. 

\begin{figure}[htb]
\vspace*{9cm}
\includegraphics{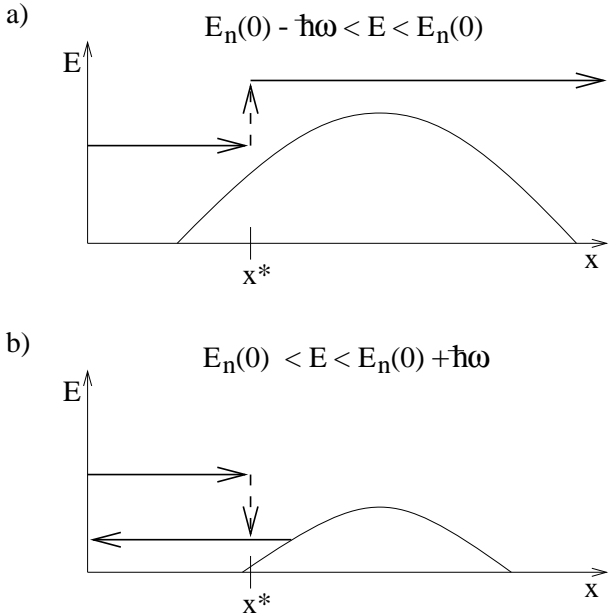}
\caption{A schematic view of the inelastic scattering processes which
give rise to induced direct current when $q<q_0$.}
\label{fig1}
\end{figure}

\begin{figure}[htb]
\vspace*{8cm}
\includegraphics{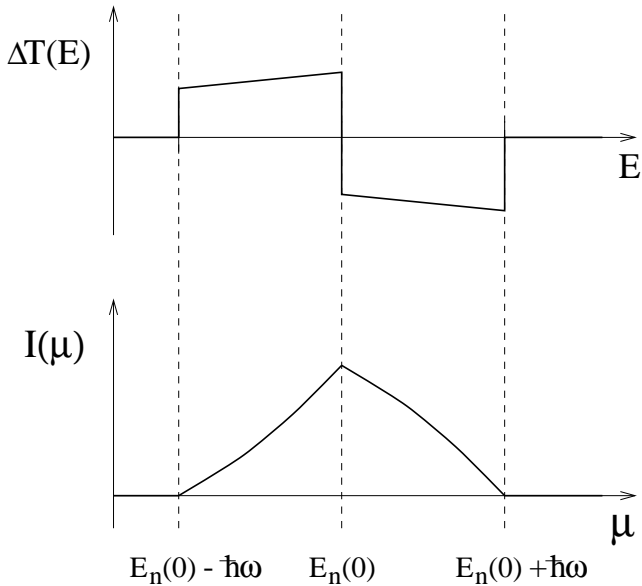}
\caption{A schematic drawing of the net transmission probability, $\Delta T(E)$,
and the induced current $I(\mu)$ when $q<q_0$.}
\label{fig2}
\end{figure}

However, the co-ordinate dependence of the channel width leads to a
finite backscattering even at $ q < q_0 $. 
If the energy $E$ of the initial state satisfies the condition 
$$E_m(0)-\hbar\omega < E < E_m(0) \, ,$$ 
then a scattering from the initial {\em reflecting} state 
(incoming from the left) with the
energy $E$ to the final  {\em propagating} (to the right) state with the energy 
$E+\hbar \omega$ is allowed.
As a result of the absorption process described above, the initial
state becomes  not fully reflecting, and the total current through the
channel {\em increases}. An emission process from the energy $E$ to the energy
$E-\hbar\omega$ will not influence the total transmission. 
On the other hand, when 
$$E_m(0) < E < E_m(0) + \hbar \omega $$
an emission process will {\em reduce} $\Delta T_{R,L}(E)$ while 
absorption is not important. Both
situations are illustrated in Fig.~\ref{fig1} (panels a and b, respectively). 
When $E > E_m(0) + \hbar \omega $, no transitions will alter the transport
properties since transitions are between states fully transmitted in the
same direction.
Fig.~\ref{fig2} is a sketch of the resulting $\Delta T(E)$ and
the DC response, $I(\mu)$. The presented qualitative  picture is
confirmed by the numerical calculations.
We do not present them in detail because the
requirement $q < q_0$ can be hardly fulfilled together with the
criterion $\hbar \omega \gg \delta E$ in realistic structures.

We believe that the effect is qualitatively similar at  $\hbar \omega
\lesssim \delta E$ (where the semi-classical approach is not valid),
but less pronounced. In fact, the DC response is very similar even
when $\hbar\omega \ll \delta E$ (as will be shown in the next section),
although the physics is quite different.

\subsubsection*{The case $q > q_0$}

For the case when $q >q _0$ things are slightly more
complicated since from equation (\ref{k2}) scattering from the 
acoustic field will now reverse the direction of propagation.
{}From  the conservation laws (\ref{k2}) one can extract the
relevant energy scale,
\begin{equation} \label{es}
\epsilon_q = \hbar^2 (q+q_0)^2/8m\, . 
\end{equation}
If $\epsilon \equiv E-E_m(0) > \epsilon_q$ then no scattering is possible
which fulfills conservation of both energy and local longitudinal momentum.
One can discriminate between the following
four energy intervals for the energy $\epsilon$.

\begin{figure}[htb]
\vspace*{16cm}
\includegraphics{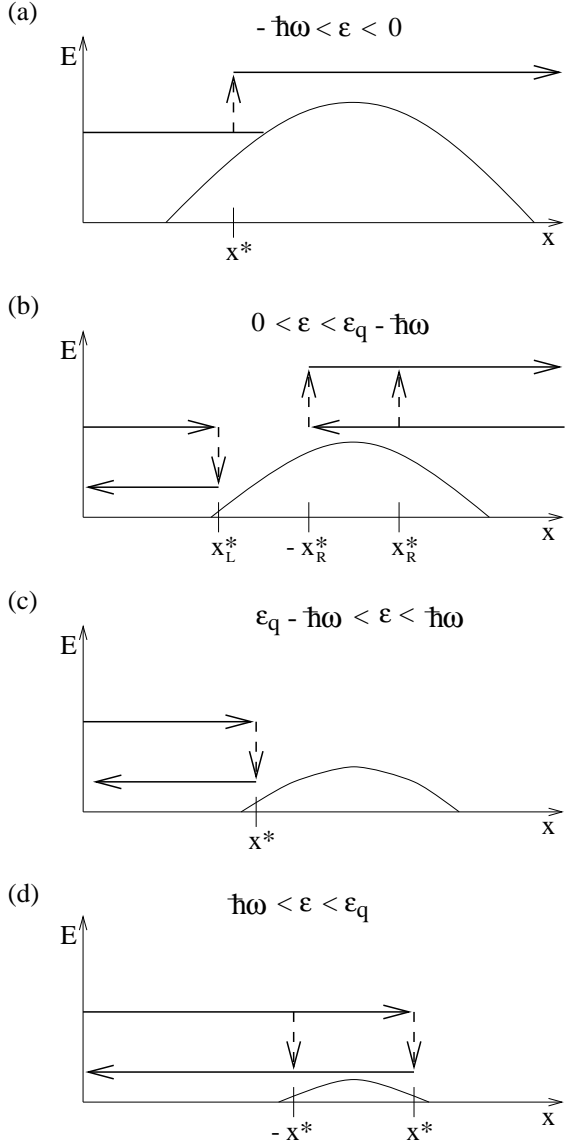}
\caption{A schematic illustration of the different scattering mechanisms
which influence the induced direct current when $q>q_0$.}
\label{fig3}
\end{figure}

In the first one (see Fig.~\ref{fig3},a),
\begin{equation} \label{i1}
- \hbar\omega < \epsilon < 0\, ,
\end{equation}
the wave with the energy $E$, incoming from the left lead and
initially reflected, absorbs
the energy $\hbar \omega$  and momentum $\hbar q$, the final state
being propagating. The process results in a positive  $\Delta T(E)$
because in this energy interval $\Delta T_{L,R}(E)$ is not affected by
inelastic scattering.

In the second region, (see Fig.~\ref{fig3},b), 
\begin{equation} \label{i2}
0 < \epsilon < \epsilon_q - \hbar\omega\, ,
\end{equation} 
the waves incoming from both the left and the right leads will
experience scattering which affects $\Delta T(E)$. The electrons, incoming from
the left,  may emit energy  $\hbar\omega$ and momentum $-\hbar q$
and then be reflected. Such a process takes place in the vicinity of
the point $x_L^*$ where the conservation laws are met.  This results
in a negative $\Delta T_{R,L}$.  
On the other hand, the particles incoming from the right may absorb an
energy quantum. It is important that there are {\em two} points,
$x^*_R$ and $-x^*_R$ where the conservation laws are met. Near both
points a  backscattering takes place, therefore the contributions to
$\Delta T_{L,R}(E)$ are also negative. 
The fact that we have {\em two} scattering points has the following
implications: 
\begin{itemize}
\item The resulting state is a superposition of an incoming wave and
{\em two} reflected waves, the reflecting points being $\pm
x^*_R$. Thus the dependence of $\Delta T_{L,R}$ vs. energy (actually, vs.
the Fermi level controlled by the gate voltage) shows  oscillations
due to interference effects between the reflected waves.

\item 
The amplitude of oscillations in $\Delta T_{L,R}(E)$ is about 
      four times greater than $\Delta T_{L,R}(E)$  for a single
      scattering event. Thus in the above mentioned region the
      interference scattering dominates $\Delta T(E)$.
\end{itemize}

In the third region (Fig.~\ref{fig3},c) 
\begin{equation} \label{i3}
\epsilon_q-\hbar\omega< \epsilon <\hbar\omega
\end{equation}
there is only one point where the scattering takes place. Consequently,
 the interference phenomena described above do not occur. Here, the
 only effect is inelastic  backscattering of the 
electrons incoming from the left via emission of a phonon.
Hence, $\Delta T(E)<0$ in this region.

The fourth region (Fig.~\ref{fig3},d) 
is defined as 
\begin{equation} \label{i4}
\hbar \omega < \epsilon < \epsilon_q \, .
\end{equation}
In this region there is still backscattering of states incoming from
the left, but now there are {\em two} scattering points giving rise
to an interference phenomena 
similar to that described above for the second region.  As a result, $\Delta
T(E)$ is negative and oscillating with a relatively large amplitude.

From the given simple analysis we observe that when $q>(3+2\sqrt{2})q_0$
the third region vanishes, and there are  interference oscillations
for all the energies in the interval
 $$0 < \epsilon < \epsilon_q .$$
Such a situation seems typical for  the phonon-drag-imaging
experiments where $\omega/2\pi$ can exceed 100 GHz.\cite{Karl} Note
that at $q \gg q_0$ the sufficient condition to employ the
semi-classical approximation is $\max \{\epsilon_q, \hbar \omega\} \gg \delta
E$ which is weaker than $\hbar \omega \gg \delta E$.  In addition, one must 
require $qL \gg 1$ in order to have any interference.
\begin{figure}[htb]
\vspace*{12cm}
\includegraphics{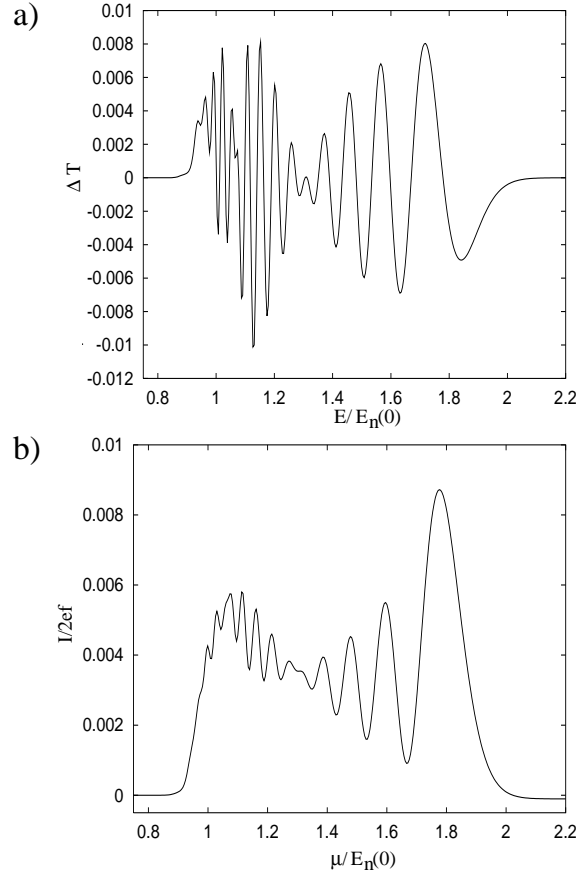}
\caption{The effect of the phonon scattering when $q=50 q_0$. 
 	The inner length of the constriction is $l\simeq 1$ $\mu$m.
	a) The induced $\Delta T(E)$.
	b) The induced dimension-less current, $I/2ef$, plotted as a
 	function of the 
	chemical potential, $\mu$. The interference oscillations are
 	clearly seen in the induced current.}
\label{fig4}
\end{figure}
Figures \ref{fig4} and \ref{fig5} show the results of numerical 
calculations for $q=50q_0$. Fig.~\ref{fig5} refers to a total
length of the constriction region of $2L$ = 2.4 $\mu$m [see 
equation~(\ref{shape})], 
which yields an inner constriction length, $l$, of approximately 1 $\mu$m.
Fig.~\ref{fig4},a shows $\Delta T(E)$ as defined in equation 
(\ref{dT_def}) while Fig.~\ref{fig5},b shows the induced dimensionless
current, $I/2ef$, as a function of the chemical potential. The interference
effects are quite pronounced, even for the induced current.
Fig.~\ref{fig5} shows the results of the numerical calculations
when $2L$ = 0.6 $\mu$m which means that the length of the inner
region is approximately $l$ = 0.25 $\mu$m. Panel a shows
the zero field  transmission. In addition, vertical lines indicating
the position of the energies $E_m(0) \pm \hbar\omega$ are
displayed. Clearly, in this case
$\hbar\omega \lesssim \delta E$. However, as it is seen in the panel
b, the interference effects in the induced current are still visible.

In an experimental situation with non-equilibrium phonons generated by
a point-like heater the phonons are not monochromatic. To observe 
interference effects in such a situation, one needs to meet the
inequality $\Delta \omega L/s  \lesssim  1$, where $\Delta \omega$ is
the characteristic width of the phonon frequency distribution. This
requirement implies $\Delta \omega /\omega \ll 1$.
\begin{figure}[htb]
\vspace*{12cm}
\includegraphics{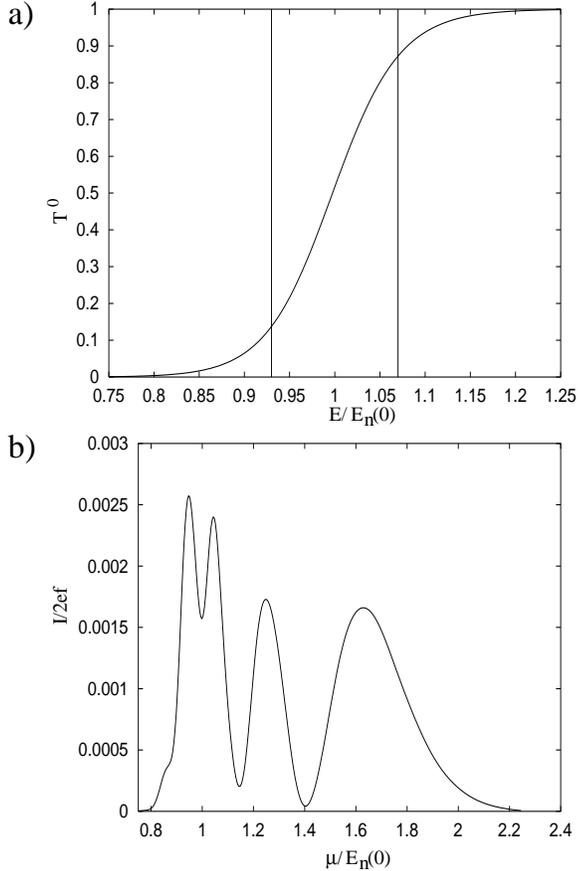} 
\caption{The effect of the phonon scattering when $q=50 q_0$. The inner 
	length of the constriction is $l\simeq 0.25$ $\mu$m.
	a) The zero field transmission, $T^0(E)$. The vertical lines indicate
	    $E_n(0) \pm \hbar\omega$. Notice that $\hbar\omega < \delta E$.
	b) The induced dimensionless current, $I/2ef$, plotted as a
	function of the 
	chemical potential, $\mu$.}
\label{fig5}
\end{figure}


\section{Low frequency regime}\label{LF}
So far, we have studied the semi-classical regime where the transition region
for $T^0(E)$, $\delta E$, is much smaller than $\hbar\omega$. However, in
many experiments using coherent SAWs\cite{Shilton1} to induce the
current,  this is not the case.  
We should therefore investigate what happens with the acoustoelectric
current at low frequencies, in particular, when  $\hbar\omega \ll \delta E$. 

\begin{figure}[htb]
\vspace*{12cm}
\includegraphics{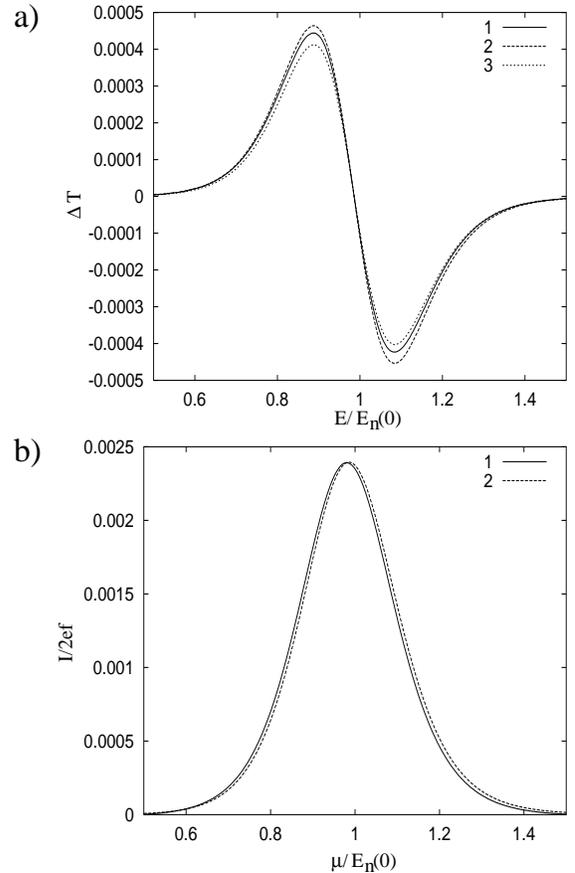}
\caption{Nearly adiabatic constriction with $2qL = 2.8$ ($q=3\times q_0$).
	a) Curve 1 shows 
	$V_0^2 \partial^2_E T^0(E)/4$, curve 2 shows $\Delta
	T_{R,L}(E)$ and curve 
	3 shows $\Delta T_{L,R}(E)$. There are no parameters fitted.
	b) Comparison of the dimensionless current, $I(\mu)/2ef$
	(line 1), and  
	$\partial_E T^0(E)\times\hbar\omega $ (line 2). Line 2 is
	scaled to match 
	the amplitude of line 1.}
\label{fig6}
\end{figure}

Suppose that the time dependence of the SAW-potential can be
considered as adiabatic in time, i.e $\omega/2\pi \ll v_F/L$ (or $qL
\ll 2 \pi v_F/s$). Let us also assume that
the SAW wavelength is large compared to the constriction length
($qL < 2\pi$). 
In the limit of infinite wavelength we must have 
$\Delta T_{R,L} = \Delta T_{L,R}$ since the symmetry $x\leftrightarrow -x$ is 
not broken any more.
To a first approximation we should then have 
\begin{eqnarray}
\Delta T_{R,L}(E) & \simeq & \Delta T_{L,R}(E) \nonumber \\
& \simeq & \frac{\omega}{2\pi}\int_0^{2\pi/\omega} T^0[E+V_0\cos
(\omega t)] \mbox{d} t - T^0(E) 
\nonumber \\
& \simeq & \frac{V_0^2}{4}\partial^2_E T^0(E).
\label{DT2}
\end{eqnarray}
It should be emphasized that in this regime there is no
distinction between the cases $q < q_0$ and $q>q_0$.
Fig.~\ref{fig6},a displays $\Delta T_{R,L}(E)$ and $\Delta T_{L,R}(E)$
together with the theoretical prediction, equation (\ref{DT2}). Note
that there are no parameters to fit. 
The results in Fig.~\ref{fig6}
were obtained with $q=3q_0$ and are also representative for the case
$q<q_0$. 

\begin{figure}[htb]
\vspace*{11.7cm}
\includegraphics{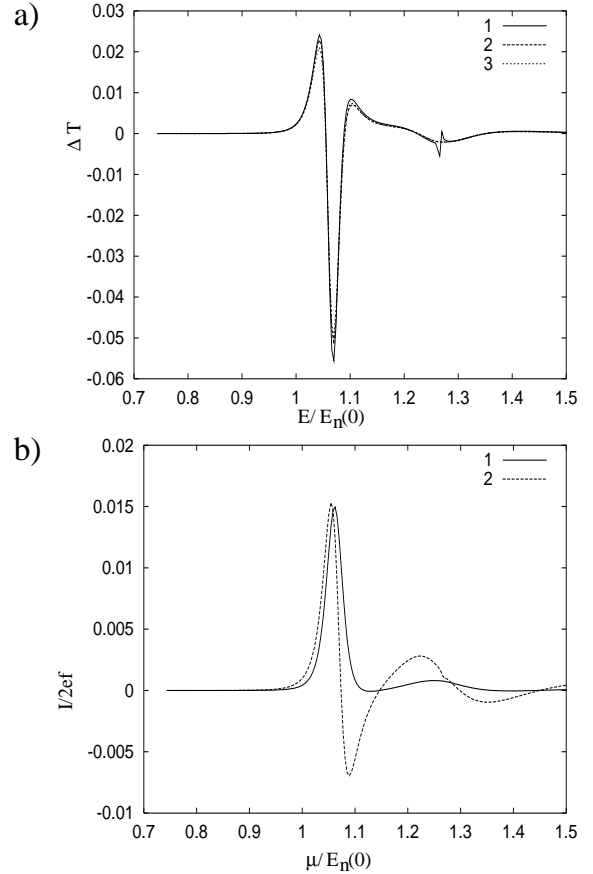}
\caption{Abrupt constriction with strong elastic scattering when $2qL = 1.12$ 
	($q=3\times q_0$). 
	a) Line 1 shows 
	$V_0^2 \partial^2_E T^0(E)/4$, line 2 shows $\Delta
	T_{R,L}(E)$ and line 
	3 shows $\Delta T_{L,R}(E)$. There are no parameters
	fitted. The kink in line 
	1 is an artifact caused by the sudden appearance of the 9th
	transverse state. 
	b) Comparison of the dimension-less current, $I(\mu)/2ef$
	(line 1), and  
	$\partial_E T^0(E)\times\hbar\omega$ (line 2). The amplitude
	of line 2 is 
	scaled to match the amplitude of line 1.}
\label{fig7}
\end{figure}

In general there is also some dependence on finite $q$, which
is the origin of the induced current. One can expect that such a
dependence at small $q$ is proportional to $q\omega$ because the effect
must be odd in $q$ and proportional to $\omega^2$ at small
frequencies. The effect of broken symmetry (finite $q$) 
can be clearly seen in Fig.~\ref{fig6},a which shows that
$|\Delta T_{R,L}(E)| \geq |\Delta T_{L,R}(E)|$. This yields the
induced dc current which is plotted in Fig.~\ref{fig7},b
together with $\hbar \omega \partial_\mu T^0(\mu)$. The curve
representing $\hbar \omega \partial_\mu T^0(\mu)$ is scaled as
to match the amplitude of the dimension-less current, $I/2ef$.
For smoothly varying constrictions, the numerical calculations
show that
\begin{equation}
\Delta T_{R,L}(E)- \Delta T_{L,R}(E)\simeq g\, \frac{V_0^2}{4s}\, 
q\omega L^2\, \partial^2_E T(E), 
\label{DT3}
\end{equation}
when $qL/2\pi < 1,  v_F/s$. The dimensional factor $g$ in
(\ref{DT3}) is  
geometry dependent and almost $L$-independent. This yields the current
\begin{equation} \label{imu}
I(\mu) \simeq g\, \frac{eV_0^2}{2sh} q\omega L^2
\partial_{\mu} T^0(\mu),
\label{current3}
\end{equation}
which explains the strong resemblance of the induced current and 
$\partial_{\mu} T^0(\mu)$. 
The similar behavior of the induced current and $\partial_{\mu}
T^0(\mu)$ agrees with the experimental findings\cite{Shilton1}. 
In an experiment, one alters the gate voltage, $V_g$,
rather than the chemical potential. Changes in $V_g$ alters both the
geometry of the constriction 
(width and length) and  the electron density within the constriction.
Since screening effects are dependent on the electron density, both the
effect of impurities and the effective coupling of the induced field to the
electrons are dependent on the electron density.
However, within small regions of the gate voltage,
the effect of changing $V_g$ should be very similar to changes in $\mu$.

\begin{figure}[htb]
\vspace*{5.8cm}
\includegraphics{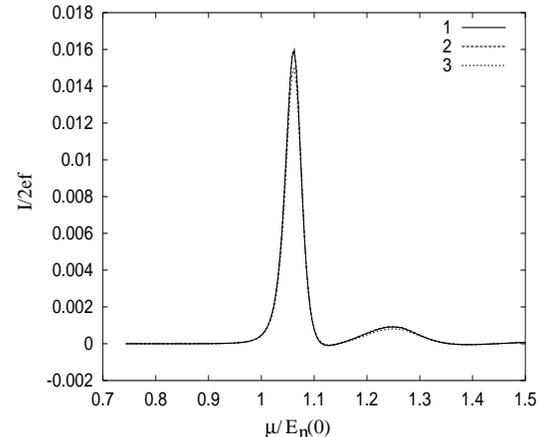}
\caption{Comparison of the dimensionless current, $I(\mu)/2ef$, times
	 $3q_0/q$ for 
	 an abrupt constriction when $q=q_0/2$ (line 1), $q=q_0
	 \sqrt{6}/2$ (line 2)  
         and $q=3 q_0$ (line 3). The figure suggests that $I \sim q\omega$.}
\label{fig8}
\end{figure}

So far we have studied the regime of weak elastic inter-mode scattering.
It is natural to ask what happens in the presence of strong elastic 
inter-mode scattering. To investigate this we have considered an
abrupt channel with
$$W(x)=\left\{ \begin{array}{lll}
W_0 &\text{at} & |x| < L/2 \, , \\
8W_0 &\text{at} & |x| > L/2 \, \end{array}\right. \, .
$$
Fig.~\ref{fig7},a shows $\Delta T_{R,L}(E)$ and $\Delta T_{L,R}(E)$
together with the result of Eq.~(\ref{DT2}). As can be seen,
Eq.~(\ref{DT2}) still holds. 
The induced current is displayed in Fig.~\ref{fig7},b together 
with $\partial_\mu T^0(\mu)$. It seems obvious that Eq.~(\ref{current3})
no longer holds, although there are some similarities between the two curves.
In particular, the scaling of the induced current, $I\sim q\omega$, is
remarkably 
well satisfied. This is shown in Fig.~\ref{fig8}. We would like to emphasize
once more that $I\sim q\omega$ and {\em not} $I\sim \omega^2$ nor
$I\sim q^2$. This fact 
has been checked explicitly by an independent choice of $q$ and $\omega$.
It is also in agreement with the recent work\cite{Totland} based on  
the Boltzmann equation. The $L^2$-dependence for the induced current
for long enough waves was also found in\cite{Totland}, according to
\cite{Blencowe} such a dependence holds also for acousto-conductivity. 
 
Eq. (\ref{DT2}) gives the acousto-conductance, $\Delta G$.
Since the numerical results agree so well with Eq.~(\ref{DT2}), it is 
tempting to speculate about the {\em second} harmonic of the ac
current. It seems most advantageous to study the second harmonic at
low enough frequencies ($qL \ll 2\pi$) 
applying a small dc bias voltage, $V_b$. In such a situation, in
addition to the dc current (\ref{imu}) and the dc Ohmic current
proportional to $V_b$, $I_0 = 2(e^2/h)\, T^0 (\mu)\, V_b$, there will
also be the ac current at frequency $2 \omega$ which is also
proportional to $V_b$. This current is originated by ac modulation of
the transparency $\Delta T (E)$.
According to Eq.~(\ref{DT2}) this current is simply given by  
\begin{equation}
I_{2 \omega} \simeq \frac{e^2 V_0^2}{2h}\cos(2\omega t) \partial^2_\mu
T^0(\mu) V_b, 
\label{2ac}
\end{equation}
Thus, the dc acousto-conductance and the ``second harmonic
conductance'', $I_{2 \omega}/V_b$, are the same.
Furthermore,
\begin{equation} \label{ratio}
\frac{I_{2 \omega}}{I_0}= \frac{\Delta
G}{G_0}\cos 2\omega t =
\frac{V_0^2}{4}\frac{\partial_\mu^2 T^0
(\mu)}{T^0 (\mu)} \cos 2 \omega t \, .
\end{equation}
Having measured the dependence $T^0 (\mu)$ from the $I-V_b$ curve, one
 can estimate the actual amplitude $V_0$ of the acoustically-induced
 potential. 
The second harmonic ac current may be also important from an
 experimental point of 
view, because the main harmonic ac signal may be masked by the
 electrical signal  from the transducer which produces SAW.\cite{end2}

\section{Summary and discussion}

 We have investigated transport properties of quantum constrictions
under the influence of a traveling time-dependent potential created either
by a coherent surface acoustic wave, or by non-equilibrium phonons. 
The results of numerical analysis based on the recursive Green function
technique are explained in the framework of a qualitative picture of
phonon-induced transitions between the electronic states.
 
The main result is that the dc {\em acoustoelectric} current induced
in the channel is crucially dependent on the (unavoidable) transitions between
propagating and reflecting states. It is those transitions that keep
the current finite at $q \le 2ms/\hbar$ in contrast with the 
prediction of the paper\cite{Gurevich} where only the transitions
between propagating states were considered. In a smooth enough channel one can
localize the important transition in space. It appears that at high
frequency, $q \ge 2ms/\hbar$, there are several regions where the
transitions take place. As a result, an interference pattern in the
dependence of the acoustoelectric current on the gate voltage is
predicted. The pattern can be observed for rather monochromatic phonon
distributions, criteria being discussed. Note that impurity potentials in
the channel can also lead to a phonon-induced backscattering and to
interference patterns.   

The set of above mentioned
results is valid for the channels with sharp steps of conductance
quantization, where the width, $\delta E$, of the energy region (where
the semi-classical approach is not valid) is less than $\hbar \omega$. 
We expect that the effect of finite temperature is the same as for
finite $\delta E$.
It is demonstrated that the acoustoelectric current exists also at
$\hbar \omega \ll \delta E$ provided  $qL \ll 2\pi$. It
is proportional $q\omega L^2$,
as well as to the derivative of the transmission probability with
respect to the Fermi level.
The role of  strong elastic scattering due to
abrupt shape of the channel is studied. It is shown that the effect
persists, and it is also proportional to $q\omega$ at low frequencies.

For low frequencies, also the acousto-conductance is studied. In particular,
when $qL \ll 2 \pi$ the explicit formula (\ref{DT2}) for the change in
transmission 
is shown to hold even in the case of strong elastic scattering. It is
speculated  
that this approach can also predict the ac current  at frequency $2 \omega$
through a biased channel.

In conclusion, we want to emphasize that the drag of electrons in
quantum constrictions by a sliding potential produced by SAWs or
non-equilibrium phonons is a rich phenomenon which involves quantum
transitions of different types. In contrast to the prediction of the
paper\cite{Blencowe} (where only the transitions between
propagating modes were considered), we observe that 
the transitions between the propagating and reflecting
states  can both {\em decrease} and {\em increase} transmission
probability and thus lead 
both to negative or positive acousto-conductance. For a detailed
comparison  with 
the experimental results one needs more information about
experimental dependencies of the effect on the frequency, temperature,
as well as on the length and the shape of the channel.

\acknowledgments
One of the authors, FAM, would like to acknowledge financial support
from Norges Forskningsr{\aa}d.

\widetext

\begin{references}

\bibitem{Shilton1} J. M. Shilton, D. R. Mace, V. I. Talyanskii, Yu. Galperin,
M. Y. Simmons, M. Pepper, and D. A. Ritchie, J. Phys.: Condens. Matter {\bf 8},
L337 (1996). 

\bibitem{Shilton2}
J. M. Shilton, V. I. Talyanskii, M. Pepper, D. A. Ritchie,
J. E. F. Frost, C. J. B. Ford, C. G. Smith and G. A. C. Jones,
J. Phys.: Condens. Matter {\bf 8}, L531 (1996).  

\bibitem{Nash96} 
G. R.Nash, S. J. Bending, Y. Kershaw, K. Eberl, P. Grambow and
K. von Klitzing, Surf. Sci.  {\bf 361/362}, 668 (1996).
 
\bibitem{Naylor} A. J. Naylor, K. R. Strickland, A. J. Kent, and M. Henini,
Surf. Sci. {\bf 361/362}, 668 (1996).

\bibitem{EG} A. L. Efros, Yu. M. Galperin,
     Phys. Rev. Lett., {\bf 64}, 1959 (1990); A. Kn\"abchen,
     Y. Levinson, and O. Entin-Wohlman, preprint cond-mat/9604137.

\bibitem{QHE1}
A. Wixforth, J. P. Kotthaus, and G. Weinman,
Phys. Rev. Lett. {\bf 56}, 2104 (1986);
A. Wixforth, J. Scriba, M. Wassermeier, J. P. Kotthaus,
G. Weinman, and W. Schlapp, Phys. Rev. {\bf B40}, 7874 (1989);
R. L. Willett, M. A. Paalanen, R. R. Ruel, K. W. West, 
L. N. Pfeiffer, and D. J. Bishop, Phys. Rev. Lett. {\bf 65}, 112 (1990);
R. L. Willett, R. R. Ruel, K. W. West, and
L. N. Pfeiffer, Phys. Rev. Lett. {\bf 71}, 3846 (1993).
\bibitem{QHE2}
A. Esslinger, A. Wixforth, R. W. Winkler,
J. P. Kotthaus, H. Nickel, W. Schlapp, and R. Losch, Solid State
Commun. {\bf 84}, 949 (1992);
J. M. Shilton, D. R. Mace, V. I. Talyanskii, M. Pepper,
M. Y. Simmons, A. C. Churchill, and D. A. Ritchie, Phys. Rev. {\bf
B51}, 14770 (1995);\ J. Phys.: Condens. Matter {\bf 7}, 7675 (1995).

\bibitem{challis} L. J. Challis, A. J. Kent, and V. W. Rampton,
Semicond. Sci. Technol. {\bf 5}, 1179 (1990); A. J. Kent, Physica B
{\bf 169}, 356 (1991); D. J. McKitterick, A. Shik, A. J. Kent, and
M. Heinini, Phys. Rev. B {\bf 49}, 2585 (1996).


\bibitem{Karl} H. Karl, W. Dietsche, A. Fischer, and K. Ploog,
Phys. Rev. Lett. {\bf 61}, 2360 (1996).

\bibitem{end0} It seems very difficult to measure linear in amplitude
effects of isolated mesoscopic systems because of small longitudinal
size. 

\bibitem{Totland} H. Totland, and Yu. Galperin, 
Phys. Rev. B {\bf 54}, 8814 (1996).

\bibitem{Gurevich} V. L. Gurevich, V. B. Pevzner, and G. J. Iafrate,
Phys. Rev. Lett {\bf 77}, 3881 (1996).

\bibitem{Gurevich1} V. L. Gurevich, V. B. Pevzner, and K. Hess,
J. Phys. Cond. Matter. {\bf 6}, 8363 (1994); Phys. Rev B {\bf 51},
5219 (1995).

\bibitem{Blencowe} M. Blencowe, and A. Shik, Phys. Rev. B {\bf 54}, 13
899 (1996). 

\bibitem{Grincwajg} A. Grincwajg, L. Y. Gorelik, V. Z. Kleiner, and R. I. Shekter,
Phys. Rev. B {\bf 52}, 12 168  (1995).
 
\bibitem{end1} In the general case, the envelope function is
proportional to the matrix element of $\exp (i {\bf q r}_\perp)$
calculated between the eigen functions $u({\bf r}_\perp)$. At small
$q_\perp$ these matrix elements yield 1. 
 
\bibitem{Datta} Supriyo Datta, and M. P. Anantram,
Phys. Rev. B {\bf 45}, 13 761 (1992).

\bibitem{Maao1} Frank A. Maa\o, and L. Y.. Gorelik, Phys. Rev. B {\bf
53}, 15 885 (1996). 

\bibitem{Landbutt} 
R. Landauer, IBM J. Res. Dev {\bf 1}, 223 (1957);
M. B\"{u}ttiker, Phys. Rev. Lett. {\bf 57}, 1761 (1986).

\bibitem{glks} L. I. Glazman, G. E. Lesovik, D. E. Khmelnitskii,
R. I. Shekhter, JETP Lett. {\bf 48}, 238 (1988). 
  
\bibitem{end2} We would like to point out that in order to investigate
the relations (\ref{DT2}) and (\ref{2ac}) one could, instead of a phonon
source, introduce an oscillating signal to the gate voltage which
induces a perturbation with $q=0, \ \omega \ne 0$.


\end{references}
\end{document}